\documentclass[nofootinbib,aps,reprint,groupedaddress]{revtex4-1}
\usepackage[T1]{fontenc}
\usepackage{graphicx}
\usepackage{color}
\usepackage{fancyhdr}
\usepackage{placeins}
\usepackage{amssymb,amsmath}
\begin{document}

%\preprint{}

\title{High resolution transient and permanent spectral hole burning in Ce$^{3+}$:Y$_2$SiO$_5$ at liquid helium temperatures}


\author{Jenny Karlsson$^{1,3}$}
\thanks{These two authors contributed equally}
\author{Adam N. Nilsson$^{1,*}$ }
\email[]{adam.nilsson@fysik.lth.se}
\author{Diana Serrano$^{1,2}$}
\author{Andreas Walther$^1$}
\author{Lars Rippe$^1$}
\author{Stefan Kr\"{o}ll$^1$}
\affiliation{$^1$ Department of Physics, Lund University, P.O. Box 118, SE-22100 Lund, Sweden}
\affiliation{$^2$ Department of Chemistry, University of Z\"{u}rich, Winterthurerstrasse 190, 8057 Z\"{u}rich, Switzerland}

\author{Philippe Goldner$^3$}
\author{Alban Ferrier$^{3,4}$}
\affiliation{$^3$ PSL Research University, Chimie ParisTech - CNRS Institut de Recherche de Chimie Paris, 75005, Paris, France}
\affiliation{$^4$ Sorbonne Universit\'es, UPMC Univ Paris 06, 75005, Paris, France}


\date{\today}

\begin{abstract}
We perform hole burning with a low drift stabilized laser within the zero phonon line of the 4f-5d transition in Ce$^{3+}$:Y$_2$SiO$_5$ at 2K. The narrowest spectral holes appear for small applied magnetic fields and are $6\pm4$ MHz wide (FWHM). This puts an upper bound on the homogeneous linewidth of the transition to $3\pm2$ MHz, which is close to lifetime limited. The spin level relaxation time is measured to $72\pm21$ ms with a magnetic field of 10 mT. 

A slow permanent hole burning mechanism is observed. If the excitation frequency is not changed the fluorescence intensity is reduced by more than 50$\%$ after a couple of minutes of continuous excitation. The spectral hole created by the permanent hole burning has a width in the tens of MHz range, which indicates that a trapping mechanism occurs via the 5d-state.
\end{abstract}

%\keywords{}

\maketitle

\section{Introduction}
\begin{figure*}
\includegraphics[width=350pt]{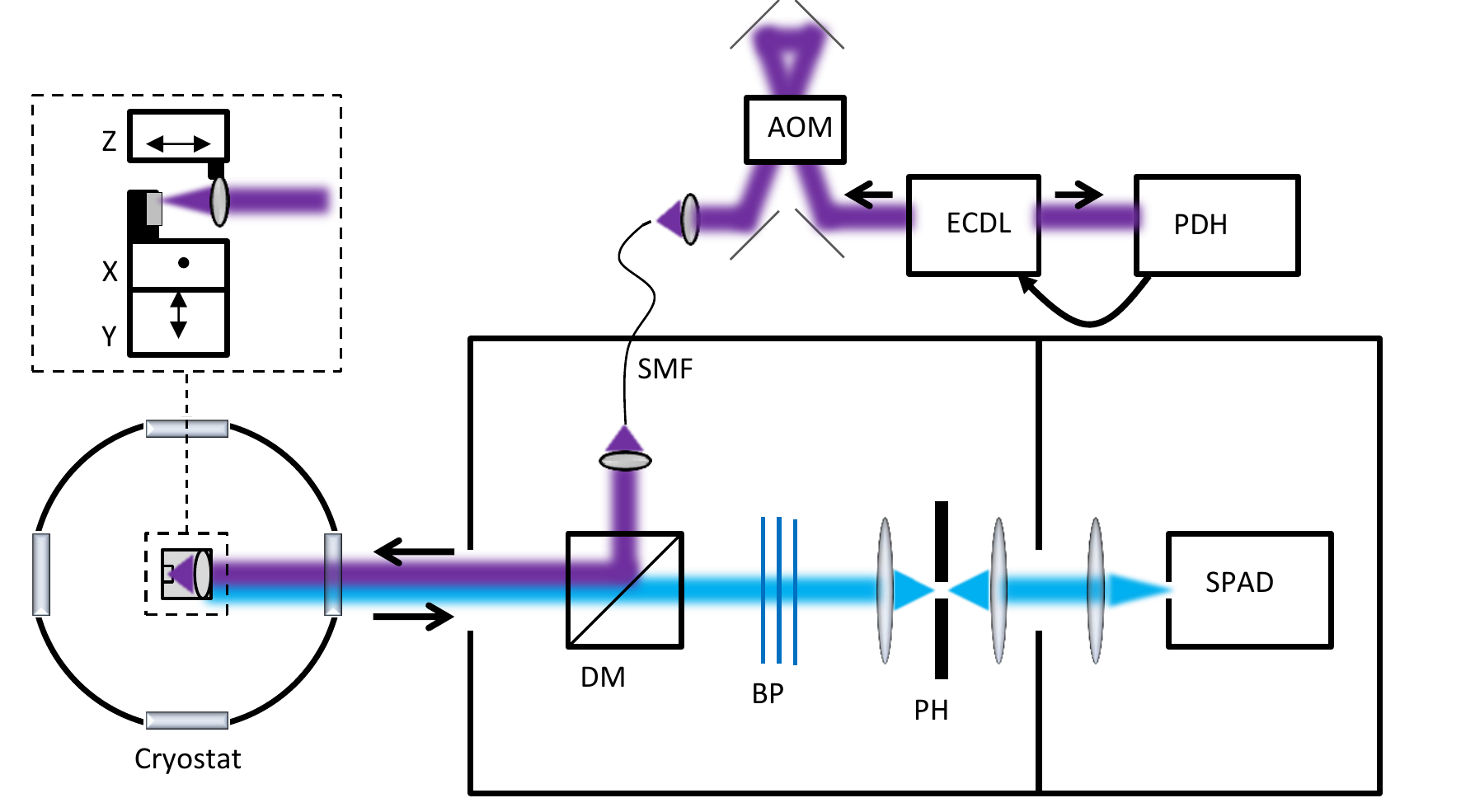} % % Important NOTE: Please make certain your figures do not include local directory paths. ex. "c:\file\sub\fig1.eps"
\caption{(Color online) Schematic of the optical setup. The purple beam represents the 371 nm laser excitation and the blue beam represents the collected fluorescence. The solid lines represents a dark enclosure. Labels are defined as follows: ECDL - External Cavity Diode Laser, PDH - Pound-Drever-Hall locking system, AOM - Acousto-Optic Modulator, DM - Dichroic Mirror, SMF - Single Mode Fiber, BP - Band Pass Filters, PH - Pinhole, SPAD - Single Photon Counting Avalanche Diode. Inside the cryostat the sample is sitting on two translators (X and Y) and the lens is attached to a third translator for focusing (Z). \label{fig:opt_setup}}%
\end{figure*}

Cerium doped yttrium orthosilicate (Ce$^{3+}$:Y$_2$SiO$_5$) crystals have been extensively studied for applications such as cathode ray tube phosphor \cite{CeYSOPhosphor1,CeYSOPhosphor2,CeYSOPhosphor3}, x-ray storage phosphor \cite{CeTrappingXRay} and as a scintillator for fast detection of x-rays and $\gamma$-rays \cite{ScintReview,ScintProp}. The 5d-state of cerium has a lifetime of about 40 ns, and gives rise to strong fluorescence in the 370-500 nm wavelength range \cite{CeYSOemission,CeFluorUVgammaRay,CeTrappingOxygen}. 

The high yield of photons is an attractive property also for detection of single ions in a crystal \cite{SingleCeStuttgart,coherentManipulation}, with possible applications in quantum information science. Currently, low concentration Ce$^{3+}$:Y$_2$SiO$_5$ is studied with the aim of implementing a quantum computing scheme where a single cerium ion is used for quantum state readout \cite{scalableDesign}. The scheme requires cerium ions to interact with the qubit ions (for example praseodymium) via the difference in permanent electric dipole moment between the ground and excited state, which has been previously measured \cite{homLine}. Depending on the quantum state of a qubit ion, the fluorescence from a single cerium ion can be switched on or off. The success of such a protocol will depend on properties of Ce$^{3+}$:Y$_2$SiO$_5$ such as the homogeneous linewidth and fluorescence yield.

The methods used for quantum information experiments are rather different from those traditionally used when studying cerium as a scintillator or phosphor. However, many of the results are related and relevant for both communities.

It is well known that $\gamma$-ray, x-ray or UV excitation of cerium ions in Y$_2$SiO$_5$ and similair crystals give rise to afterglow, a very long lived luminescence emission that can last for hours \cite{CeTrappingXRay,CeTrappingOxygen,CeTrapping80K,CeTrappingLowTemp,AfterglowLSO}. The cause of afterglow is a slow recombination emission from electron and hole traps in the crystal lattice. The trapping of charge carriers worsens the performance of Ce$^{3+}$:Y$_2$SiO$_5$ as a scintillator \cite{ScintReview}, but enables its application as a storage phosphor \cite{CeTrappingXRay}. Charge traps can be detrimental for single ion detection, since when an electron belonging to a cerium ion gets trapped it might not recombine for hours and the ion is lost from view. Many mechanisms that can lead to charge trapping have been suggested, but exactly which mechanisms that are present in the crystal under certain conditions remains to be clarified.

Trapping mechanisms in Ce$^{3+}$:Y$_2$SiO$_5$ and undoped Y$_2$SiO$_5$ have been previously studied by means of thermally stimulated luminescence (TSL) \cite{TSLCeYSO,CeTrappingXRay,defectsCeYSO,unboundO,CeTrapping80K}, electron spin resonance (ESR) \cite{defectsCeYSO,unboundO} as well as absorption, excitation and emission spectroscopy \cite{CeTrappingXRay,CeTrappingOxygen,defectsCeYSO,CeTrappingLowTemp}, with excitation by x-rays \cite{CeTrappingXRay,unboundO}, $\gamma$-rays \cite{TSLCeYSO} or UV light \cite{CeTrappingXRay,CeTrappingOxygen,defectsCeYSO,CeTrapping80K,CeTrappingLowTemp}. These methods provide information about trap depths, the concentration of some defects in the crystal and possible charge trapping and recombination mechanisms.

In this work, hole burning and charge trapping in a Ce$^{3+}$:Y$_2$SiO$_5$-crystal at a temperature of 2K is investigated. Contrary to most previous experiments a low drift stabilized laser targeting the zero phonon line (ZPL) of the cerium 4f-5d transition is used. The high spectral resolution allows for new conclusions to be drawn regarding trapping and hole burning mechanisms. 

When applying a magnetic field, redistribution of ions in the spin levels of the ground state within the ZPL gives rise to a spectral hole with a lifetime of $72\pm21$ ms. The width of the spectral hole for the smallest fields is measured to $6\pm4$ MHz, which puts an upper limit on the homogeneous linewidth of the transition to $3\pm2$ MHz and confirms that the linewidth is close to lifetime limited \cite{homLine}.

A slower and much longer lived hole burning mechanism is observed as a decrease of the cerium fluorescence signal under continuous excitation for several minutes. The created spectral hole has a lifetime of hours. A similar observation was previously made in Ce$^{3+}$:LuPO$_4$, Ce$^{3+}$:YPO$_4$ \cite{holeBurningLuPOandYPO}, and Ce:YAG \cite{Xia2015}. To the best of our knowledge it is the first time a persistent trapping mechanism in Ce$^{3+}$:Y$_2$SiO$_5$ has been observed with high spectral resolution within the ZPL. The width of the created spectral hole is in the tens of MHz range. The high frequency selectivity shows that trapping occurs via the excited 5d-state of cerium. A rate equation model that is able to capture the characteristic decrease of the fluorescence signal, is put forth.

\section{Experimental setup}

A sketch of the optical setup can be seen in figure \ref{fig:opt_setup}. An external cavity diode laser (ECDL) centered at 371 nm was used to excite the cerium ions and perform hole burning. The frequency of the ECDL was stabilized to a low drift cavity made of ultra low expansion glass (ULE) using the Pound-Drever-Hall technique \cite{ZhaoThesis}. An acousto-optic modulator (AOM) in double pass configuration allowed the laser frequency to be continuously scanned over a range of 200 MHz. After the AOM, the laser beam was sent through a single mode fiber to obtain a clean Gaussian TEM$_{00}$-mode before sending it to the sample. A maximum power of about 200 $\mu$W reached the sample. The polarization directly after the laser was linear, but when reaching the sample the polarization was probably slightly elliptical after passing several optical components.

To collect fluorescence from the sample a home built fluorescence detection setup was used \cite{microscope}. The laser beam was reflected off a dichroic mirror with high reflectivity at 371 nm, and high transmission in the range 385-450 nm corresponding to the fluorescence emission from Ce$^{3+}$:Y$_2$SiO$_5$. The laser beam was sent into a liquid helium bath cryostat (Oxford Instruments Spectromag) where the sample was kept at a temperature of 2K. The laser beam was focused about 100 $\mu$m below the surface of the sample by a small lens with a numerical aperture of 0.85 mounted inside the sample space of the cryostat. Both the sample and the lens were mounted on nanometer precision translators (Attocube ANP51RES). The laser focus was estimated to have a diameter of 1 $\mu$m in the crystal. The saturation intensity of the 4f-5d transition corresponds to a laser power of 14 $\mu$W \cite{homLine}.

The fluorescence was collected in the backward direction by the lens inside the cryostat. The dichroic mirror and three interference filters were used to block unwanted laser light. The three interference filters together transmits <0.01$\%$ at the laser wavelength of 371 nm, and transmits $>$ 90$\%$ only in the range 385-425 nm to select fluorescence from cerium in site 1 while blocking most of the fluorescence from cerium ions in site 2 \cite{CeFluorescence}. A 50 mm lens was used to focus the fluorescence through a 25 $\mu$m pinhole which blocks light generated out of focus in the crystal. The fluorescence was measured using a single photon counting avalanche diode (countBlue-250B). 

\section{Sample}

\begin{figure}
\includegraphics[width=220pt]{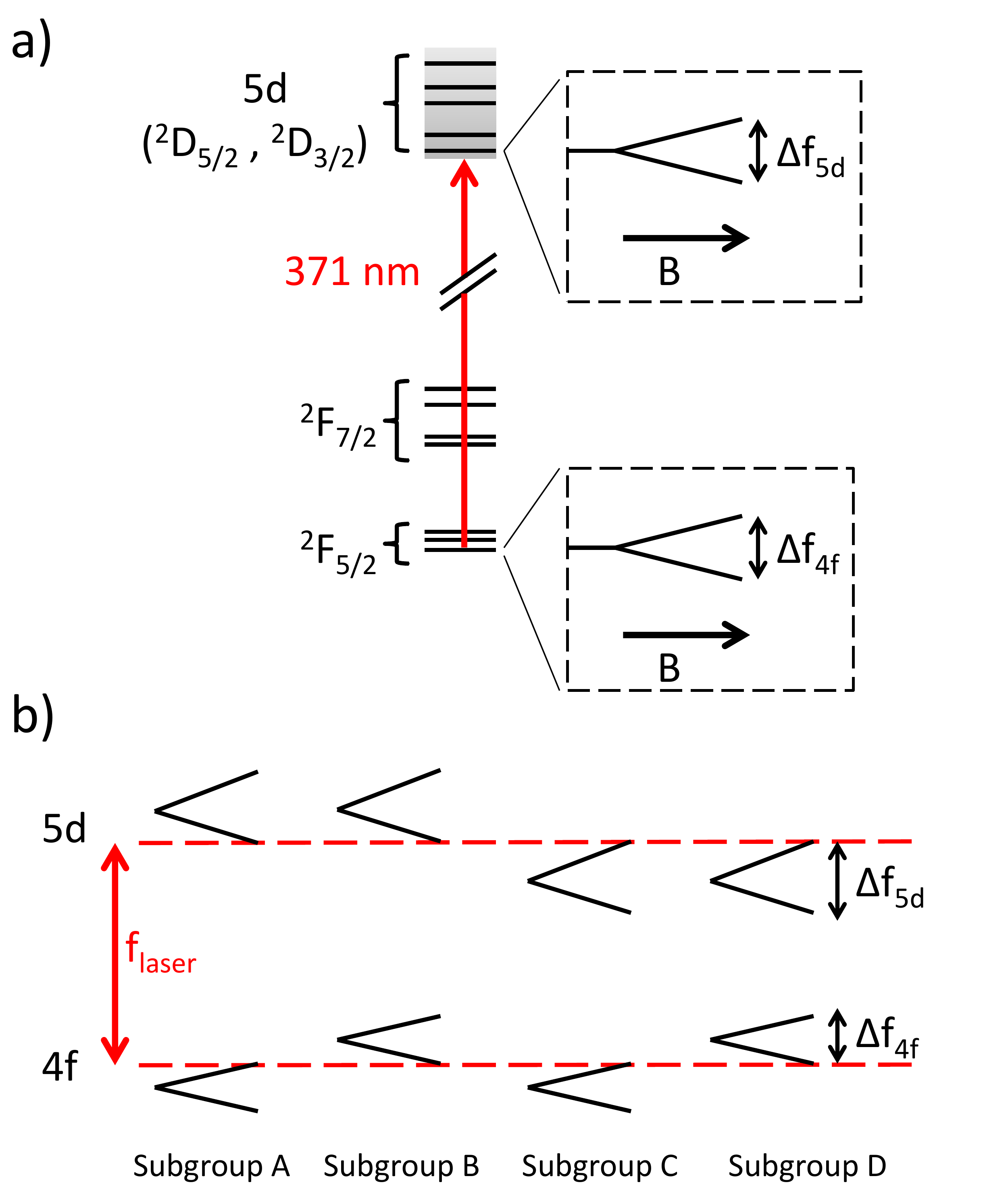}%
\caption{(Color online) a) The energy levels of Ce$^{3+}$:Y$_2$SiO$_5$. When applying a small magnetic field the lowest stark level of the ground state will split up into two Zeeman levels separated by $\Delta f_{4f}$, and the lowest level of the excited state will split up into two levels separated by $\Delta f_{5d}$. b) Because of the inhomogeneous broadening of the absorption line of Ce$^{3+}$:Y$_2$SiO$_5$, four different groups of ions absorb at each laser frequency $f_{laser}$ when a small magnetic field is applied. \label{fig:inhom}}
\end{figure}

The sample used in the following experiments is a very low concentration Ce$^{3+}$:Y$_2$SiO$_5$ crystal grown by the Czochralski method. The concentration of cerium ions relative to yttrium is in the order of 10$^{-7}$. The crystal is  kept at 2K in a liquid helium bath cryostat throughout all measurements.

Ce$^{3+}$ has a very simple energy level structure with two electronic configurations, 4f and 5d, below the conduction band of the crystal. The 4f-5d transition is electric dipole allowed and gives rise to strong fluorescence. The 4f ground state is split into two fine structure multiplets, which are further split into crystal field levels, all of which are doubly degenerate at zero magnetic field. The energy level structure of Ce$^{3+}$:Y$_2$SiO$_5$ (site 1) is shown in figure \ref{fig:inhom} a). In this work the 4f-5d transition always refers to the transition between the lowest energy crystal field level in the 4f and 5d states respectively.

Cerium ions can occupy two crystal sites with different oxygen coordination \cite{twoCrystalSites}. In this work excitation of cerium takes place within the zero phonon line of ions in site 1, at $371$ nm, with an inhomogeneous linewidth of 37 GHz. The detected fluorescence in the selected range 385-425 nm originates almost exclusively from cerium ions in site 1, with a signal to background ratio of 100 in the center of the line. The background is in this case measured by tuning the laser towards longer wavelengths 1 nm away from the zero phonon line of cerium in site 1, and consists mainly of overlapping fluorescence from cerium ions in crystal site 2.

When applying an external magnetic field the doubly degenerate fine structure levels will split into two Zeeman levels. In each of the two yttrium sites, the cerium ions can occupy two nonequivalent magnetic sites \cite{CeYSO_gFactor_spinRel}. For a magnetic field applied along the b-axis of the crystal the Zeeman splittings of the two magnetic sites overlap. This is the case in the present study \cite{CeYSO_ESR}.

\section{Hole burning in a magnetic field}

\begin{figure}
\includegraphics[width=240pt]{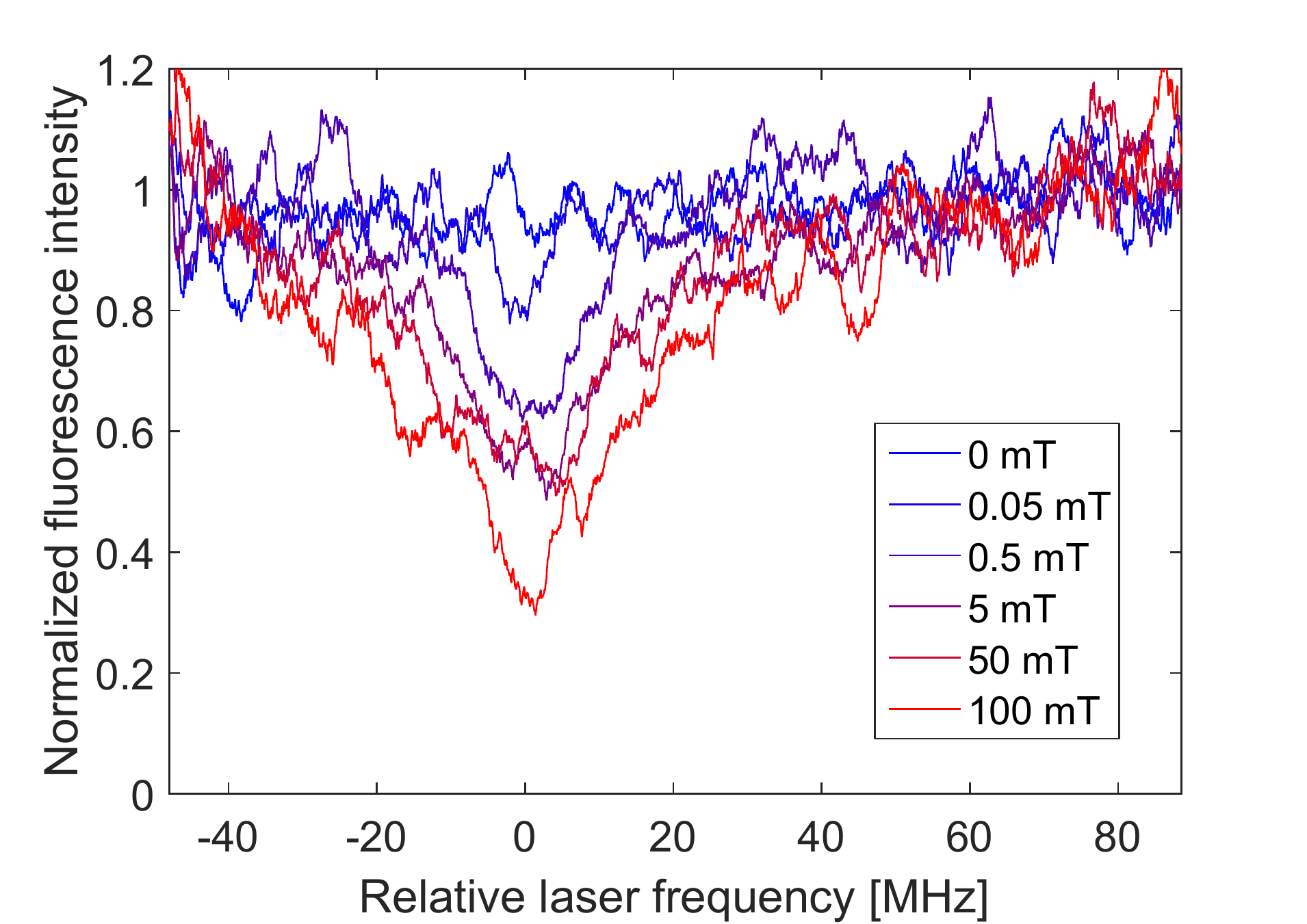}%
\caption{(Color online) A spectral hole is burnt in the inhomogeneous absorption profile by redistribution of ions in the Zeeman levels. Spectral holes created with different magnetic fields are shown in different colors, and the field strength is given in the inset. The data has been smoothed with a moving average over 100 points, corresponding to around 4 MHz on the frequency axis. Although the incoming power was monitored to be the same, the overall fluorescence intensity was monotonically decreasing for the different measurements with a maximum decrease of around $30 \%$. This is thought to be due to a beam alignment drift causing a reduction in the efficiency of the detection setup. Therefore, the signals are normalized so that the average signal level above 50 MHz is set to 1. \label{fig:fieldStrength}}
\end{figure}

\begin{figure}
\includegraphics[width=240pt]{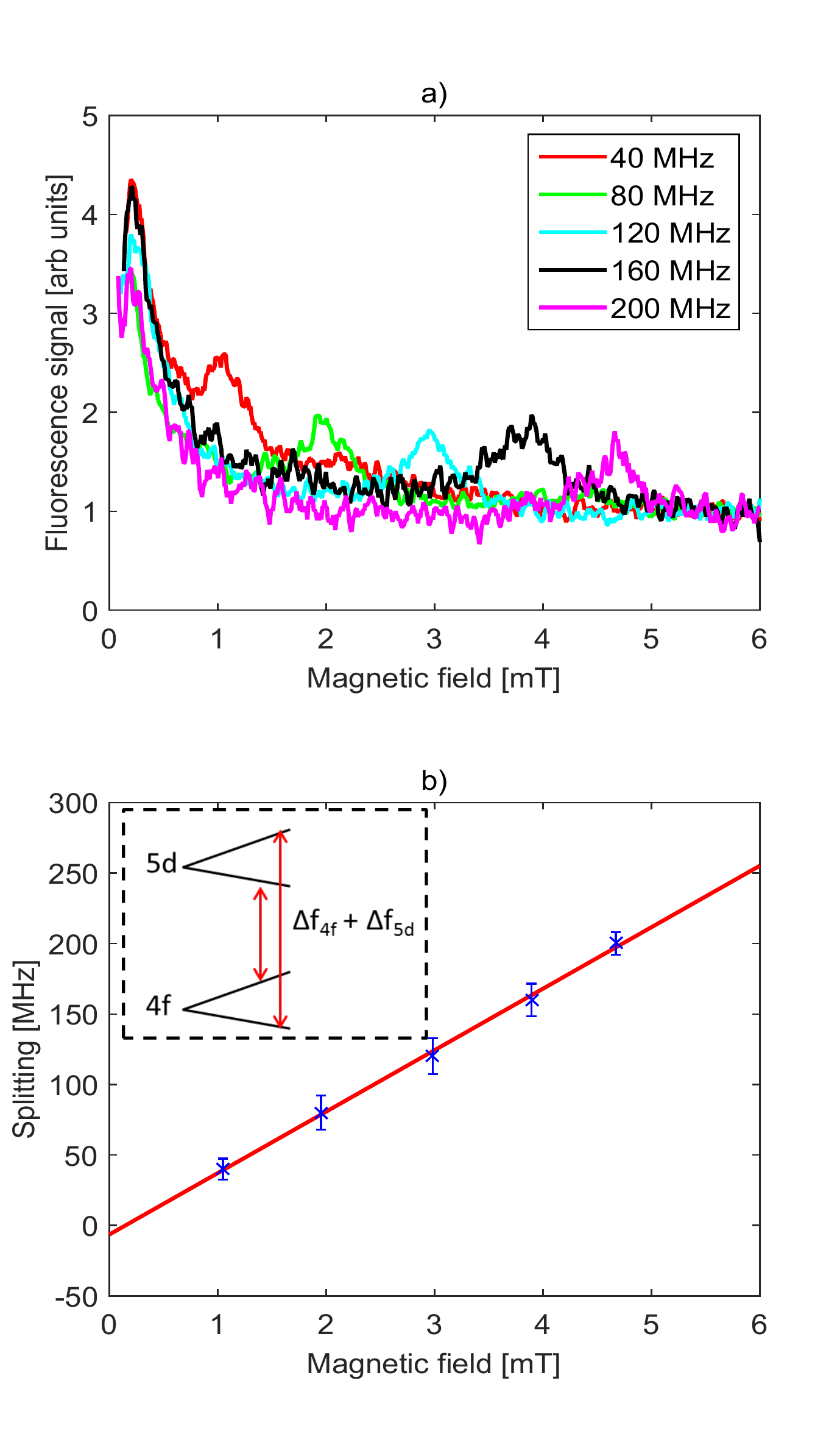}%
\caption{(Color online) The Zeeman splitting of the ground state and the excited state was measured by rapidly switching the laser between two frequencies separated by $\Delta f_{laser}$ while scanning the magnetic field. a) When $\Delta f_{laser}$ matches the Zeeman splitting of the ground state or the sum of the ground state and excited state splitting, a resonance peak is detected. $\Delta f_{laser}$ for the different curves is shown in the inset. The data has been smoothed with a moving average over 7 points, corresponding to a magnetic field difference of $66 \text{ } \mu$T. Note that the peaks at $0.2$ mT corresponds to a cancellation of a stray magnetic field, for more details see the main text. b) The measured sum of the ground and excited state level splitting as a function of magnetic field with a linear fit. \label{fig:splitting}}
\end{figure}

\begin{figure}
\includegraphics[width=240pt]{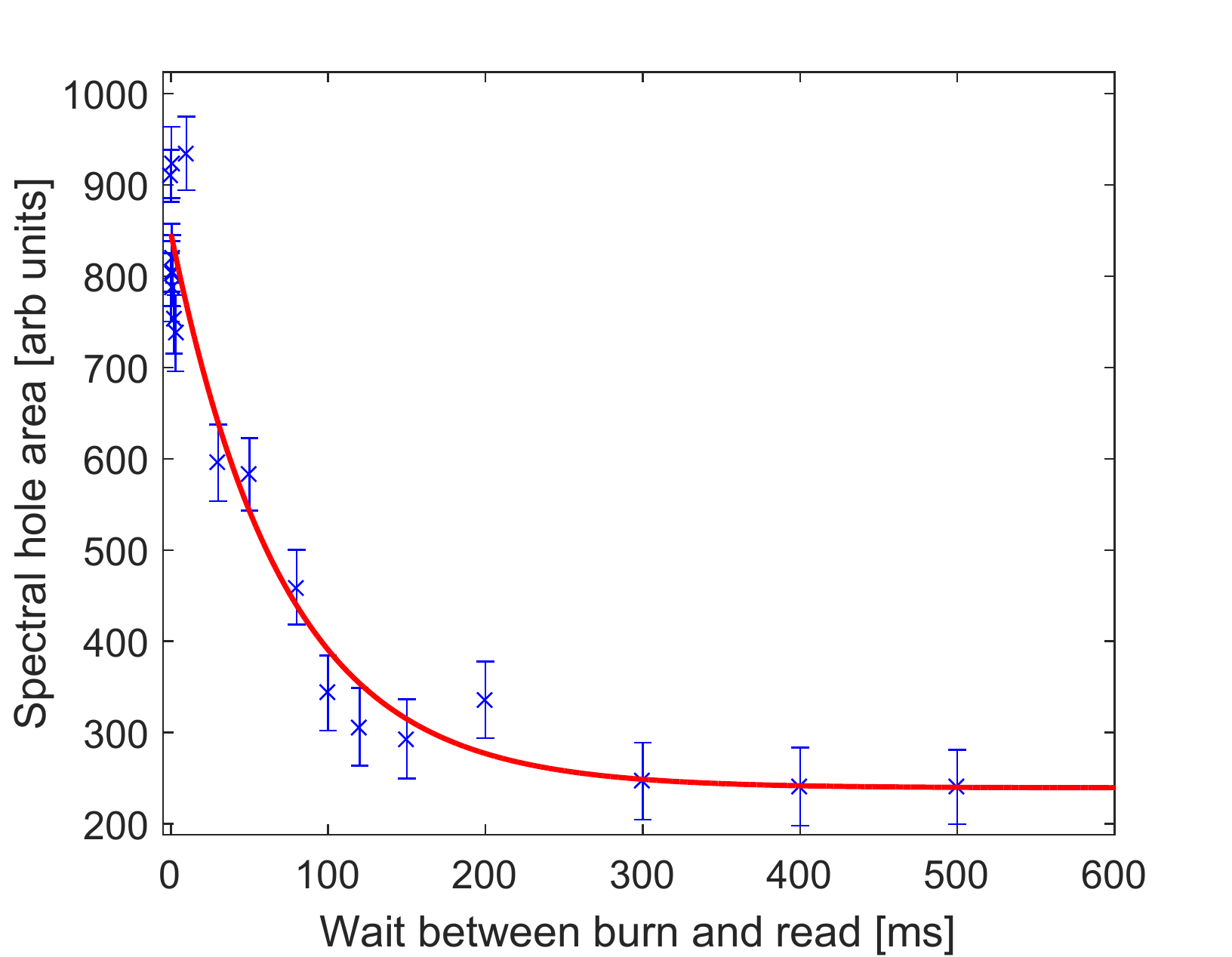}%
\caption{(Color online) The decay time of the spectral hole in the Zeeman levels is $72\pm21$ ms. The error bars show the standard deviation of the data and the red curve is an exponential fit. \label{fig:holeDecay}}
\end{figure}

A small external magnetic field (0.05-100 mT) was applied to the crystal in the direction parallel to the laser beam, to split the doubly degenerate fine structure levels. A laser pulse of 300 $\mu$s duration and 20 $\mu$W of power ($\approx$ 1.4 times the saturation intensity) was applied close to the center of the inhomogeneous absorption profile to create a spectral hole. The hole was detected by scanning the laser frequency over a 200 MHz interval in 100 $\mu$s while recording the fluorescence from the sample. A wait time allowed the system to relax to thermal equilibrium. The experiment was repeated about 1000 times to average data. 

As can be seen in figure \ref{fig:fieldStrength}, a spectral hole is created for an applied field as small as 0.05 mT. With increasing magnetic field the spectral hole gets deeper, and wider.

The FWHM of the spectral holes was measured by fitting a constant minus a Lorentzian curve to the data. More details about the data treatment can be found in the Supplementary Material.  For a field of 0.05 mT the FWHM of the Lorentzian fitted to the spectral hole is $6\pm4$ MHz, which puts an upper limit on the 4f-5d homogeneous linewidth of half that value \cite{Moerner1988}, $3\pm2$ MHz. 

Because of the inhomogeneous broadening of the zero phonon line, a fixed laser frequency $f_{laser}$ simultaneously drives all four different transitions from the two Zeeman ground levels to the two excited Zeeman levels, but for different ions, as shown in figure \ref{fig:inhom} b). The ground state splitting, $\Delta f_{4f}$, can be calculated using the known g-factor of Ce$^{3+}$:Y$_2$SiO$_5$ \cite{CeYSO_gFactor_spinRel,CeYSO_ESR}. For a field along the b-axis, as in the present measurement, the splitting is 19 MHz/mT. 

To measure the Zeeman splitting in our sample, the AOM in the laser beam was switched between two frequencies, separated by an amount $\Delta f_{laser}$, every 4 $\mu$s. If the two laser frequencies matches one transition from each ground state in the same subgroup of cerium ions, as shown in figure \ref{fig:inhom} b), hole burning will be less efficient since the ions will be repumped every 4 $\mu$s. This happens when $\Delta f_{laser}$ is equal to the ground state Zeeman splitting $\Delta f_{4f}$, when it is equal to the sum of the ground state and the excited state splitting, $\Delta f_{4f}$ + $\Delta f_{5d}$ (subgroup A and D), or when it is equal to the difference between the ground state and the excited state splitting, $\Delta f_{4f}$ - $\Delta f_{5d}$ (subgroup B and C). 

To find these points, $\Delta f_{laser}$ was kept fixed while the magnetic field was scanned from 0 to 6 mT, see figure \ref{fig:splitting} a). A strong resonance peak, and for small $\Delta f_{laser}$ also a weaker peak at about twice the magnetic field, could be observed. The weaker peaks show a magnetic field dependence of 17.9$\pm$1.5 MHz/mT (80$\%$ confidence interval) which agrees well with to the known ground state Zeeman splitting of 19 MHz/mT. The stronger peaks are at lower magnetic fields and should therefore correspond to the sum of the ground state and the excited state splitting, $\Delta f_{4f}$ + $\Delta f_{5d}$, and can be determined from figure \ref{fig:splitting} a). See figure \ref{fig:splitting} b) for a linear fit to the data, with a slope of 43.4$\pm$1.7 MHz/mT, showing that the excited state splitting is similar to the ground state splitting with a value of 25.5$\pm$3.2 MHz/mT.

The peak corresponding to $\Delta f_{4f}$ - $\Delta f_{5d}$ is at much higher magnetic fields and can not be seen in figure \ref{fig:splitting}.

In figure \ref{fig:splitting} a) a strong fluorescence peak around 0.2 mT can also be seen. This peak does not change with $\Delta f_{laser}$ and thus has nothing to do with repumping caused by switching between the two laser frequencies. It was first assumed that the peak corresponds to partial cancellation of the earth's magnetic field, and hence to an overlap of the Zeeman levels. The field strength of 0.2 mT is however too large to be attributed to the earth's magnetic field alone, which is around 50 $\mu$T at the lab location. As the direction of the applied magnetic field was reversed the peak at 0.2 mT did not show up, and all resonance peaks were shifted by a small amount. This indicates that the peak is the actual point of zero field. A measurement using a magnetic field probe inserted into the sample space of the cryostat confirms that there is a stray magnetic field of around 0.2 mT inside the cryostat, even when the current through the coils is zero. The origin of this magnetic field has not yet been investigated.

The lifetime of the spectral hole was measured by fixing the magnetic field at 10 mT and varying the wait time between the burn pulse and the scanned readout of the spectral hole. The burn pulse was in this case 1 ms long, with 30 $\mu$W of laser power. The spectral hole decays to 1/e of the initial area in $72\pm21$ ms (80$\%$ confidence interval), see figure \ref{fig:holeDecay}. This is in the same order as the previously reported spin-lattice relaxation time of Ce:Y$_2$SiO$_5$ \cite{CeYSO_gFactor_spinRel}.

It is interesting to note that even 500 ms after the burn pulse a small spectral hole remained. This persistent spectral hole can be related to another trapping mechanism, for example ionization of Ce$^{3+}$ into Ce$^{4+}$. This will be discussed further in the next section. 

\section{Permanent trapping}
\label{sec:PermanentTrapping}

\begin{figure}
\includegraphics[width=\linewidth]{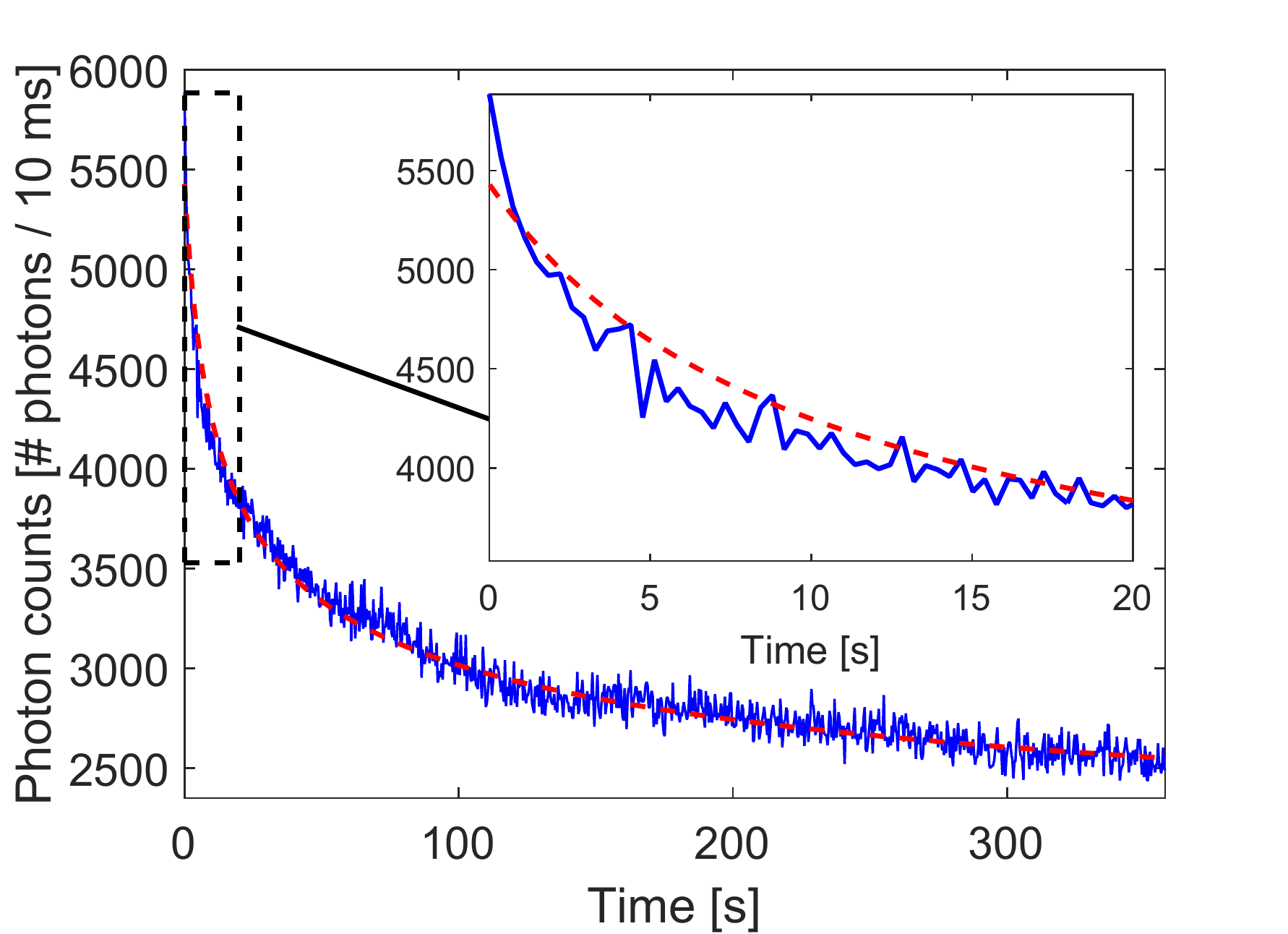}
\caption{(Color online) At time zero the crystal is translated so that the laser focus interacts with a new set of ions. The laser excitation is kept at a constant power of 20 $\mu$W. The fluorescence signal slowly goes down over a couple of minutes to around 50$\%$ of the initial signal strength (blue line). The data was taken at an applied magnetic field of 0.2 mT, where the hole burning due to redistribution in the Zeeman levels has a minimum. Other sets of data were taken at zero applied field and show the same behavior. The simulation result of the rate equation model explained in the main text is seen in the red dashed line.\label{fig:slowBurn}}
\end{figure}

When exciting the cerium ions continuously for a few minutes it was noted that the fluorescence signal slowly decreases, see figure \ref{fig:slowBurn}. Initially the signal drops fast, within a few seconds the signal has decreased by about 25$\%$. After the initial quick drop the fluorescence slowly decreases over a time scale of minutes, indicating that the ions are slowly transferred to a long lived trap state. The fluorescence signal drops to slightly below 50 $\%$ of the initial value. The laser was turned off for up to 40 minutes and then turned on again without any sign of recovery of the signal. Thus the lifetime of the trap state is at least hours. Very similar data was measured in Ce:YAG by Xia et al. \cite{Xia2015}. 

Using the AOM in the laser beam, a sequence of frequency jumps was done with varying magnitudes from 10 to 100 MHz. Immediately after a sufficiently large frequency jump the fluorescence signal is back and a new decay starts. The increase of the fluorescence signal versus frequency jump was used to estimate the spectral hole width. The estimate gave a spectral hole width (FWHM) around 70 MHz. This shows that the trapping mechanism is frequency selective on the tens of MHz scale. Thus trapping has to happen via the 5d-state of the cerium ions, since the zero phonon line of the 4f-5d transition is the only transition with narrow homogeneous linewidth which can allow hole burning on the tens of MHz scale. The reason for the increased frequency width, compared to the few MHz linewidth of the 5d transition, is however unknown. 

\begin{figure}
\includegraphics[width=150pt]{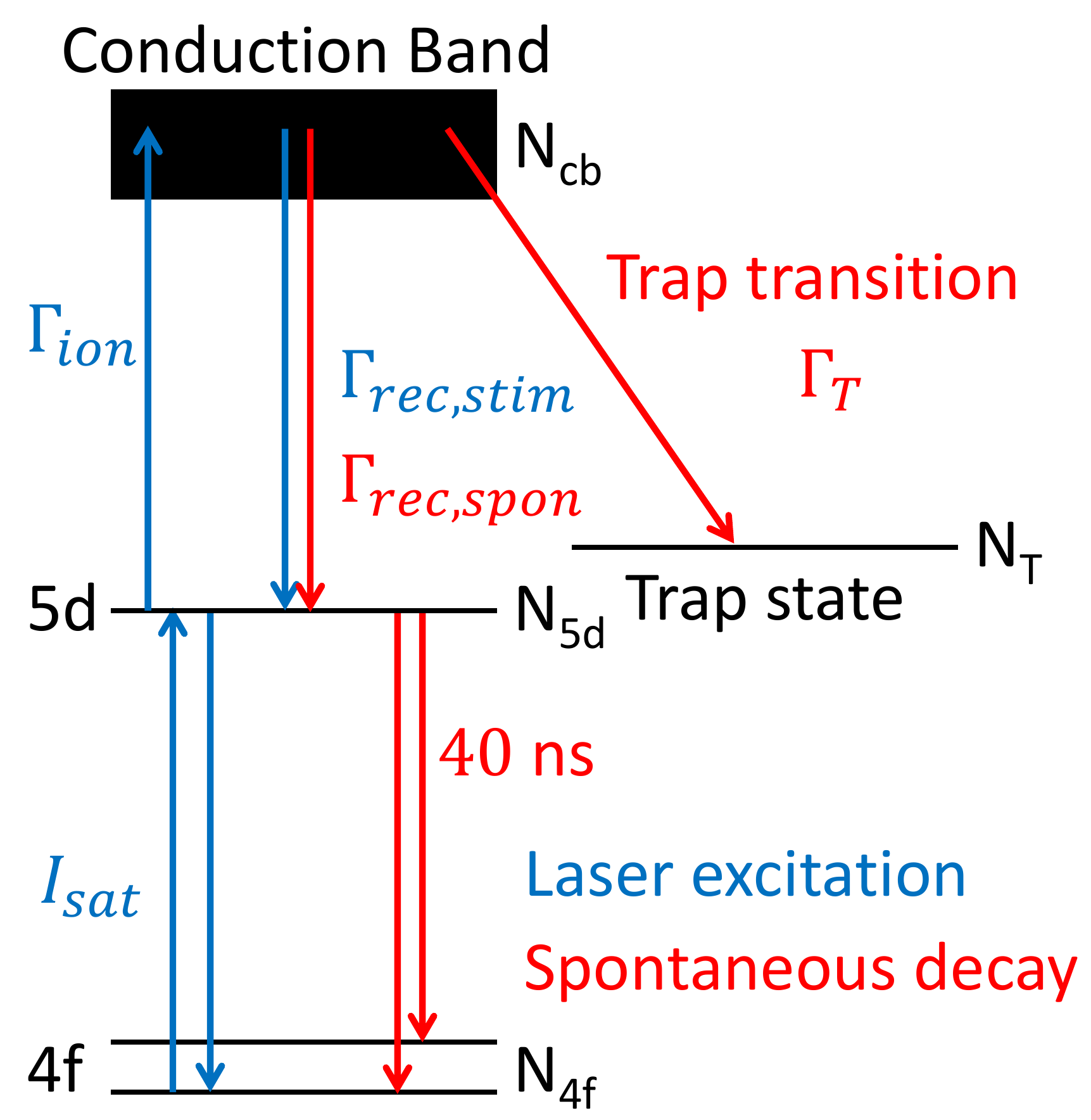}%
\caption{(Color online) A simple level diagram showing a model of the permanent trapping. The 4f-5d transition will reach steady state on the 100 ns time scale. To simplify the model, it is also assumed that the 5d-conduction band transition will be in a steady state. Trapping can occur by spontaneous decay from the conduction band to the permanent trapping state. \label{fig:levels}}
\end{figure}

To investigate this permanent hole burning mechanism further, a rate equation model using the 4f, 5d, conduction band, and trap state seen in figure \ref{fig:levels} was analyzed \cite{Loudyi2007, Xia2015}. In the model, an electron reaching the conduction band can either escape to a permanent trap via spontaneous decay, or recombining with the Ce ion. To simplify calculations, the first three states (4f, 5d and conduction band) are, at each time instance, assumed to be in a steady state that depends on the excitation and saturation intensities. The 4f-5d saturation intensity has previously been measured \cite{homLine} to be $1.4 \cdot 10^7$ W/m$^2$. Furthermore, the model assumes that cerium ions are continuously distributed both in space and frequency, and that fluorescence is emitted in a continuous manner from the 5d excited state. 

Since the detection setup uses a pinhole for increased spatial resolution in the focus direction, the simulation includes a spatially depending collection efficiency of the emitted fluorescence. Integration over all three spatial dimensions, as well as over the frequency distribution of the cerium ions to account for off-resonant excitation and spectral broadening due to saturation, are performed and the result of the simulations is a time depending detected signal. 

Finally, we take into account that all cerium ions do not seem to get trapped (the fluorescence does not tend to zero in figure \ref{fig:slowBurn}), and the model accounts for this by adding a constant fluorescence to the signal. Note that this means that the trap state should not be an internal state of the cerium ion. It is possible that the electron in the conduction band finds another crystal impurity, for example an oxygen vacancy \cite{Loudyi2007}. 

The results of the simulation can be seen in figure \ref{fig:slowBurn} as the dashed red curve. Since several parameters needed for the simulation were unknown, an optimization of three parameters was performed to get the best fitting. The first parameter was the trap rate. The other two parameters were the fraction of cerium ions that cannot be trapped, and an overall signal normalization constant (to account for unknown collection efficiency etc.). Furthermore, the cross section for ionization to the conduction band and the cross section for recombining of the electron from the conduction band with the 5d excited state were estimated by the results from Pavlov et al. \cite{Pavlov2013}, who estimated these for Ce:LiYF$_4$ and Ce:LiLuF$_4$. Some parameters used in the simulation can be seen in table \ref{table:param}.

\begin{table}
  \begin{tabular}{|  l | r |} 
    \hline
    Rate for ionization, $\Gamma_{ion} = \Gamma_{rec,stim}$   						& $3\cdot 10^{4} \text{ s}^{-1}$ 				\\ \hline
    Rate for recombination, $\Gamma_{rec, spon}$						& $2\cdot 10^{8} \text{ s}^{-1}$ 				\\ \hline
    Rate for permanent trapping, $\Gamma_T$ 					& $7\cdot 10^{4} \text{ s}^{-1}$				\\
    \hline
  \end{tabular}
\caption{The ionization and recombination rates are estimated from the Ce:LiYF$_4$ and Ce:LiLuF$_4$ results in Pavlov et al. \cite{Pavlov2013} and assumed to be similar for Ce:Y$_2$SiO$_5$. The permanent trapping rate was optimized in the simulation to best fit the data in figure \ref{fig:slowBurn}. See Supplementary Materials section \ref{sec:Model} for more details on the simulation model and parameter estimations. }
\label{table:param}
\end{table}

For a more thorough mathematical explanation of the model, see the Supplementary Materials section \ref{sec:Model}. 

It should be noted, that this is only one physically reasonable model, and not the only such model. 

The intensity dependence of the permanent trapping was also investigated and fluorescence curves for seven different laser powers was experimentally measured. These, together with simulation fits, can be found in the Supplementary Material section \ref{sec:Intensity}.

\section{Conclusion}
Hole burning within the 4f-5d zero phonon line of cerium in site 1 in a very low concentration Ce$^{3+}$:Y$_2$SiO$_5$ crystal was performed with MHz spectral resolution. With a small magnetic field applied, redistribution of population in the spin levels give rise to a spectral hole. The width of the created spectral hole shows that the homogeneous linewidth of the 4f-5d transition is around $3\pm2$ MHz, which is close to lifetime limited. The lifetime of the spectral hole was measured to be $72\pm21$ ms, with a magnetic field of $10$ mT. 

A very slow hole burning mechanism was discovered which decreases the fluorescence signal to about 50$\%$ in a few minutes of continuous excitation. The fact that the fluorescence intensity does not tend to zero suggests that not all cerium ions in this crystal can become trapped. The created spectral hole has a width in the tens of MHz range and a lifetime of at least hours. To model the the permanent hole-burning, a rate equation model where cerium ions could get excited from the 5d state to the conduction band and from there spontaneously decay to permanent traps was introduced, and the results fitted well with the experimental data. 

One possible explanation for this permanent trapping is that an electron in the conduction band finds an impurity in the crystal, for example an oxygen vacancy \cite{Loudyi2007}, which traps an electron and leave behind a Ce$^{4+}$-ion. However, further studies are needed to draw final conclusions on the exact trapping mechanism. 

Permanent trapping can be a serious problem for detection of single cerium ions in Y$_2$SiO$_5$. The time scale on which trapping happens is however long, many seconds, and data suggests that not all cerium ions get trapped. Work by Kornher et al. \cite{Kornher2016}, also suggests that annealing crystals under Ar + H$_2$ atmosphere makes them photostable. For these reasons it seems like single ion detection is still fully possible. 

If the permanent trapping mechanism can be reversed, which several studies suggests \cite{Loudyi2007, Xia2015, Kornher2016}, so that the cerium fluorescence signal comes back on demand, it might have interesting uses e.g. for long-lived spectrally tailored filters and slow light applications \cite{Beavan2013, Zhang2012, Sabooni2013}.

\begin{acknowledgments}
This work was supported by the Swedish Research Council (VR), the Knut and Alice Wallenberg Foundation (KAW), (Marie Curie Action) REA Grant Agreement No. 287252 (CIPRIS), Lund Laser Center (LLC), and the Nanometer Structure Consortium at Lund University (nanoLund).
\end{acknowledgments}

\section{Supplementary Material}
\renewcommand{\thefigure}{S\arabic{figure}}

\setcounter{figure}{0}

\subsection{Data treatment of spectral holes}
To read out a spectral hole the laser was scanned over a 200 MHz interval with a double pass AOM. The fluorescence intensity was recorded as a function of laser frequency. The hole burning and readout scan was repeated about 1000 times to collect data. Since the AOM efficiency depends on the driving frequency, the power of the laser incident on the sample changed during the scan. A reference detector was used to monitor the laser power and during each scan a calibration curve was recorded together with the fluorescence intensity. 

\begin{figure*}
\includegraphics[width=\linewidth]{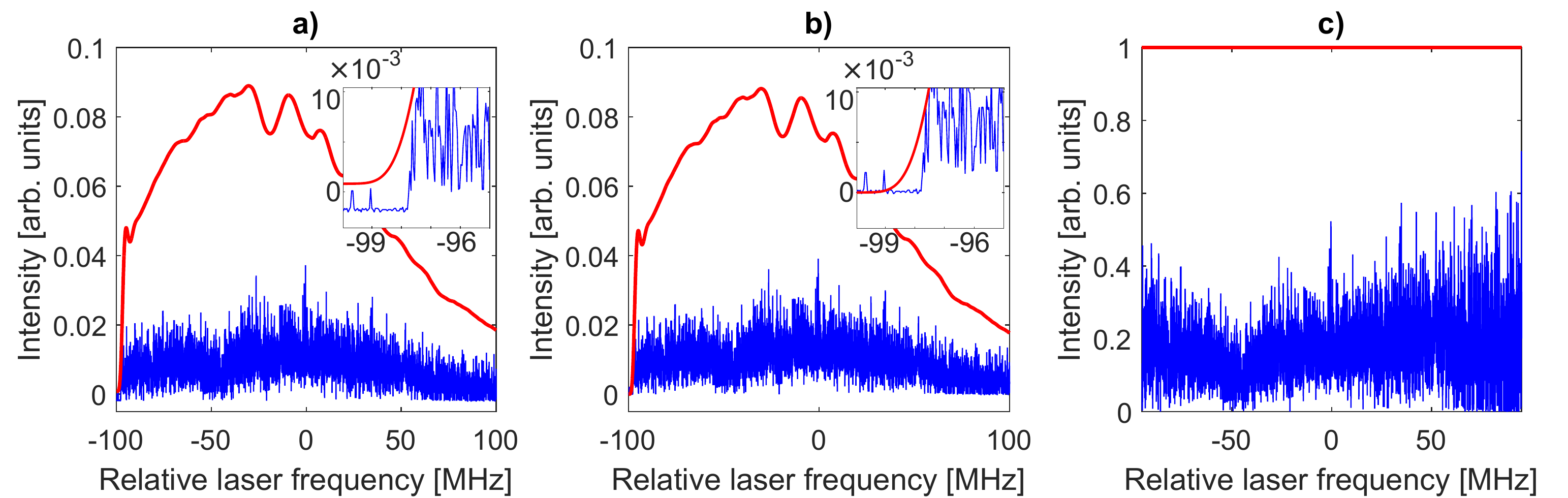}%
\caption{(Color online) The figures show a frequency scan over a spectral hole located at -50 MHz. In all figures the blue trace is fluorescence and the red trace is the laser power. The data is treated in two steps. a) Raw data with an offset which can be seen in the leftmost points of the curves (see inset for zoomed in view). b) The offset has been subtracted from both curves (see inset for zoomed in view). c) The fluorescence curve has been divided by the laser power point by point. \label{fig:dataTreat_hole}}
\end{figure*}

Data treatment was done in two steps, illustrated in figure \ref{fig:dataTreat_hole}. A part of each saved curve contains a small region where the AOM was turned off, which shows the detector background counts for both the laser power monitor and the fluorescence detector. This background was subtracted from the signals (figure \ref{fig:dataTreat_hole} b)). The fluorescence curve was divided by the laser power curve to obtain a straight baseline on the vertical axis (figure \ref{fig:dataTreat_hole} c)). This method was used for both the data shown in figure 3 and figure 5 in the main article.

In figure 3 of the main article the data was also smoothed by a moving average over 100 points, corresponding to a frequency of around $4$ MHz, to illustrate the spectral holes. In figure 4 a) of the main article the data was smoothed by a moving average over 7 points, corresponding to $66$ $\mu$T. Otherwise no smoothing was applied to the data.

\subsection{Error bars for spectral holes}
By taking a data trace without burning a spectral hole, the RMS error for each data point ($\sigma_{point}$) can be estimated. This is done by normalizing the data as shown in \ref{fig:dataTreat_hole}, and then calculate the RMS of the data relative to a baseline at y = average of signal where there is no hole. Data points where the laser intensity is zero are excluded.

The spectral hole area is a sum over all points of the data curve, and the error is calculated as a sum of squares of the point RMS error, as in equation \ref{sumOfSquares}. This is the case in figure 5 of the main article.

\begin{equation}
\sigma_{area} = \sqrt{\sum{\sigma_{point}^2}}
\label{sumOfSquares}
\end{equation}

To estimate the spectral hole depth and width a constant minus a Lorentzian curve was fitted to the data. No smoothing was applied since this might influence the results. The lowest FWHM estimated was $6\pm4$ MHz for an applied magnetic field of 0.05 mT.

\subsection{Permanent trapping model} \label{sec:Model}
In this section the permanent trapping model briefly described in section \ref{sec:PermanentTrapping} of the main text will be further explained. The model consists of four levels, 4f, 5d, conduction band, and a trap state, seen in figure \ref{fig:levels} in the main text, where the first three levels are assumed to always be at a steady state. The transition between 4f and 5d have a saturation intensity of $I_{sat} = 1.4 \cdot 10^7$ $\text{W}/\text{m}^2$ that has previously been measured \cite{homLine}. The relation between the population $N_{4f}(t)$ in the 4f state and the population $N_{5d}(t)$ in the 5d state is therefore;

\begin{equation}\label{eq:pop_4f-5d}
\begin{split}
	\frac{N_{4f}(t)- N_{5d}(t)}{N_{4f}(t)+N_{5d}(t)} &= \frac{1}{1+ \frac{I_{exc}^{4f-5d}}{I_{sat}}}			\\
	\frac{N_{4f}(t)}{N_{5d}(t)} &= 1 + \frac{2\cdot I_{sat}}{I_{exc}^{4f-5d}} \equiv R_1
\end{split}
\end{equation}

where $I_{exc}^{4f-5d}$ is the excitation intensity for the 4f-5d transition. An expression for $I_{exc}^{4f-5d}$ will be given later in this section, but the model does not include any local field corrections. 

From measurements performed on Ce:LiYF$_4$ and Ce:LiLuF$_4$ in Pavlov et al. \cite{Pavlov2013} the cross section for ionization and recombining between the 5d and conduction band levels are around $\sigma_{ion} = 10^{-18}\text{ cm}^2$ and $\sigma_{rec} = 10^{-16} \text{ cm}^2$. Furthermore, the frequency Full-Width-at-Half-Maximum (FWHM) of the excited state photoionization spectra is around $\Delta f = 82 $ THz. We here assume that similar values can be used for Ce:YSO. 

The rate of transitions $\Gamma_{ion}$ for stimulated ionization or $\Gamma_{rec, stim}$ stimulated recombination from the conduction band is calculated below \cite{Hilborn1982}; 

\begin{equation}\label{eq:pop_5d-cb}
\begin{split}
	\Gamma_{ion} = \Gamma_{rec, stim} &= \sigma_{ion} \frac{I_{exc}^{5d-cb}}{E_{\gamma}} 				\\
	E_{\gamma} &= \frac{h c_0}{\lambda_0}
\end{split}
\end{equation}

where $I_{exc}^{5d-cb}$ is the excitation intensity for the 5d-conduction band transition, $E_{\gamma}$ is the photon energy, $h$ is Planck's constant, $c_0$ the speed of light in vacuum, and $\lambda_0=371$ nm is the vacuum excitation wavelength. This gives a rate of transition of $\Gamma_{ion} = 3 \cdot 10^4$ s$^{-1}$. 

If $\Gamma_{rec}$ for spontaneous recombination from the conduction band to 5d is assumed to be radiative with a deexcitation wavelength of $\lambda_{deex} = 371$ nm, it can be calculated using the following equation \cite{Hilborn1982}; 

\begin{equation}
\begin{split}
	\Gamma_{rec, spon} &= \frac{4 \sigma_0}{\lambda_{deex}^2} \cdot \frac{g_{5d}}{g_{cb}}		\\
	\sigma_0 &= \int_{-\infty}^{\infty} \sigma_{rec} d\omega \approx \sigma_{rec} \cdot 2\pi \cdot \Delta f
\end{split}
\end{equation}

If we assume that $g_{5d} = g_{cb}$ and use $\sigma_{rec} = 10^{-16} \text{ cm}^2$ and $\Delta f = 82 $ THz, the spontaneous transition rate from the conduction band is $\Gamma_{rec, spon} = 2 \cdot 10^8$ s$^{-1}$. Note that the model only requires a rate of the spontaneous decay from the conduction band and that the mechanism of the decay is not important. We have therefore used a radiative decay assumption even though a decay by phonon might be more probable. Using these rates we can now put up a steady state solution between the 5d and the conduction band populations; 

\begin{equation}
\begin{split}
	N_{5d}(t) \cdot \Gamma_{ion} &= N_{cb}(t) \cdot (\Gamma_{rec, stim} + \Gamma_{rec, spon})		\\
	\frac{N_{5d}(t)}{N_{cb}(t)} &= 1 + \frac{\Gamma_{rec, spon}}{\Gamma_{ion}} \equiv R_2
\end{split}
\end{equation} 

The total number of cerium ions can be estimated to be roughly $N = 6 \cdot 10^{10}$ ions/m$^3$/Hz, for a relative cerium to yttrium concentration of $10^{-7}$ and an inhomogeneous profile of $30$ GHz. 

Given that the total number of cerium ions, $N$, should remain constant the following equation can be put forth and modified to solve for the population in the conduction band as a function of time;

\begin{equation}
\begin{split}
	& N_{4f}(t) + N_{5d}(t) + N_{cb}(t) + N_T(t) = N 									\\
	& (R_1 \cdot R_2 + R_2 + 1) \cdot N_{cb}(t) = N - N_T(t) 								\\
	N_{cb}(t) &= \frac{1}{(R_1 \cdot R_2 + R_2 + 1) } \cdot (N-N_T(t))  \equiv 					\\
	&\equiv k \cdot (N-N_T(t)) 				
\end{split}
\end{equation}

The rate equation for the trap state can now be solved; 

\begin{equation}
\begin{split}
	\frac{dN_T}{dt} &= \Gamma_{trap} \cdot N_{cb}(t) = 								\\
	&= \Gamma_{trap} \cdot k \cdot (N-N_T(t))										\\
	\Rightarrow N_T(t) &= N \cdot (1-e^{-\Gamma_{trap} \cdot k \cdot t}) 
\end{split}
\end{equation}

where we have used that at time $t = 0$ no cerium ions are trapped, i.e., $N_T(0) = 0$. Given this expression for $N_T(t)$ we can write the population in $N_{5d}(t)$ as; 

\begin{equation}
\begin{split}
	N_{5d}(t) &= R_2 \cdot k \cdot (N-N_T(t)) =										\\
	&= R_2 \cdot k \cdot N \cdot e^{-\Gamma_{trap}\cdot k \cdot t}
\end{split}
\end{equation}

We now assume that the fluorescence emitted from the ions are proportional to the population in $N_{5d}(t)$ with a rate of $f_0 = 1/(40 \text{ ns})$, since the lifetime of the 5d state is $40$ ns. 

\begin{equation}\label{eq:flour}
	f(t) = f_0 \cdot N_{5d}(t) 
\end{equation}

So far we have only included the time dependence on this fluorescence signal $f(t)$, but the excitation intensity $I_{exc}^{4f-5d / 5d-cb}$ depends spatially on where in the focus the cerium ion is excited, i.e.; 

\begin{equation}\label{eq:int_spatial}
\begin{split}
	I_{exc}^{4f-5d / 5d-cb}(r, \theta, z) &= I_0 \cdot \left(\frac{w_0}{w(z)} \right)^2 \cdot e^{-\frac{2r^2}{w(z)^2} }		\\
	I_0 &= \frac{2P_0}{\pi w_0^2}															\\
	w(z) &= w_0 \sqrt{1 + \left(\frac{z}{z_r}\right)^2}												\\
	w_0 &= \frac{\text{FWHM}}{\sqrt{2\cdot\ln{2}}}													\\
	z_R &= \frac{\pi w_0^2}{\lambda_0/n}
\end{split}
\end{equation}

where $r$, $\theta$ and $z$ are cylindrical coordinates, $w_0$ is the beam waist given by the FWHM of the setup (estimated to be $\text{FWHM}=1 \text{ } \mu$m), $P_0$ is the power of the incoming laser light, $z_R$ is the Rayleigh length, and $n = 1.8$ is the refractive index of the crystal. Furthermore, not all cerium ions are resonant with the incoming laser light due to the inhomogeneous broadening. It is therefore necessary to include a detuning, $\Delta$, dependence on the amount of fluorescence emitted from the ions. Here it is assumed that only the 4f - 5d transition is narrow enough to be considered, with a linewidth of $\Gamma_{hom}^0 = 4$ MHz. However, the transition might be power broadened which gives an increased linewidth according to $\Gamma_{hom} = \Gamma_{hom}^0 \cdot \sqrt{1 + \frac{I_{exc}^{4f-5d}(r, \theta, z)}{I_{sat}} }$ \cite{Siegman1986}. This detuning dependence will affect the excitation intensity between the 4f and 5d states as follows: 

\begin{equation}\label{eq:I_4f-5d}
	I_{exc}^{4f-5d}(r, \theta, z, \Delta) = I_{exc}^{4f-5d}(r, \theta, z) \cdot \frac{(\Gamma_{hom}/2)^2}{\Delta^2 + (\Gamma_{hom}/2)^2}	
\end{equation}

The collection efficiency, $coll$, also has a spatial dependence (same as in equation \ref{eq:int_spatial});

\begin{equation}
	coll(r, \theta, z) = coll_0 \cdot \left(\frac{w_0}{w(z)} \right)^2 \cdot e^{-\frac{2r^2}{w(z)^2} }
\end{equation}

where $coll_0 = 1.6 \%$ \cite{microscope}. 

We can now write up the equation for the total detected signal as follows;

\begin{equation}
	%S(t) = \int_{r=r_{min}}^{r_{max}} \int_{\theta=0}^{2\pi} \int_{z=z_{min}}^{z_{max}} \int_{\Delta=\Delta_{min}}^{\Delta_{max}} f(t, r, \theta, z, \Delta) \cdot coll(r, \theta, z) \cdot r \text{ } dr d\theta dz d\Delta
	S(t) = \int f(t, r, \theta, z, \Delta) \cdot coll(r, \theta, z) \cdot r \text{ } dr d\theta dz d\Delta
\end{equation}

where $f(t, r, \theta, z, \Delta)$ is given by equation \ref{eq:flour} using equation \ref{eq:I_4f-5d} and \ref{eq:int_spatial} to introduce the spatial and detuning dependencies in equations \ref{eq:pop_4f-5d} and \ref{eq:pop_5d-cb}, respectively. The integration limits were chosen in such a manner that beyond the limits any impact on the detected signal is negligible. The limits used were; $r=0\rightarrow 4 \text{ } \mu$m, $\theta = 0\rightarrow 2\pi$, $z=-60\rightarrow 60 \text{ } \mu$m, and $\Delta=-100\rightarrow 100$ MHz.

To compare the simulation signal to the experimental data signal, it needs to be rescaled. It should also include a constant fluorescence background from the cerium ions that cannot be trapped, which is assumed to be linearly dependent on the incoming laser power. The scaling was done as follows;

\begin{equation}
	S_{scaled}(t) = A \cdot S(t) + B \cdot P_0
\end{equation}

where $A$ and $B$ are the scaling and background / laser power parameters, respectively. 

Given the incoming laser power, $P_0$, for a given experiment, the only parameters missing from the model are $\Gamma_{trap}$, $A$, and $B$. A MatLab algorithm (fminsearch) was used to calculate the best estimate for these parameters using a least square optimization to estimate the error between the simulation results and the experimental data. Using the experimental data seen in figures \ref{fig:2uW}-\ref{fig:44uW} in the Supplementary Material, the algorithm globally optimized $\Gamma_{trap} \approx 15 \cdot 10^4 \text{ s}^{-1}$ and $B \approx 9.2\cdot 10^{6}$ detected counts / W, individually optimizing the parameter $A$ to each individual signal to account for fluctuations in the detection efficiency between measurements. $A$ varied between $\approx 0.007 \rightarrow 0.013$. When collecting this data the collection efficiency was much lower than optimal, which is why the detected number of photons for similar incoming laser power is much less than in the data seen in figure 6 in the main text. Running the optimization on the data from figure 6 in the main text alone gives the following values; $\Gamma_{trap} \approx 7 \cdot 10^4 \text{ s}^{-1}$, $B \approx 9.4\cdot 10^{7}$ detected counts / W, and $A \approx 0.19$. 

\subsection{Permanent trapping intensity dependence} \label{sec:Intensity}
\begin{figure}
\includegraphics[width=0.48\textwidth]{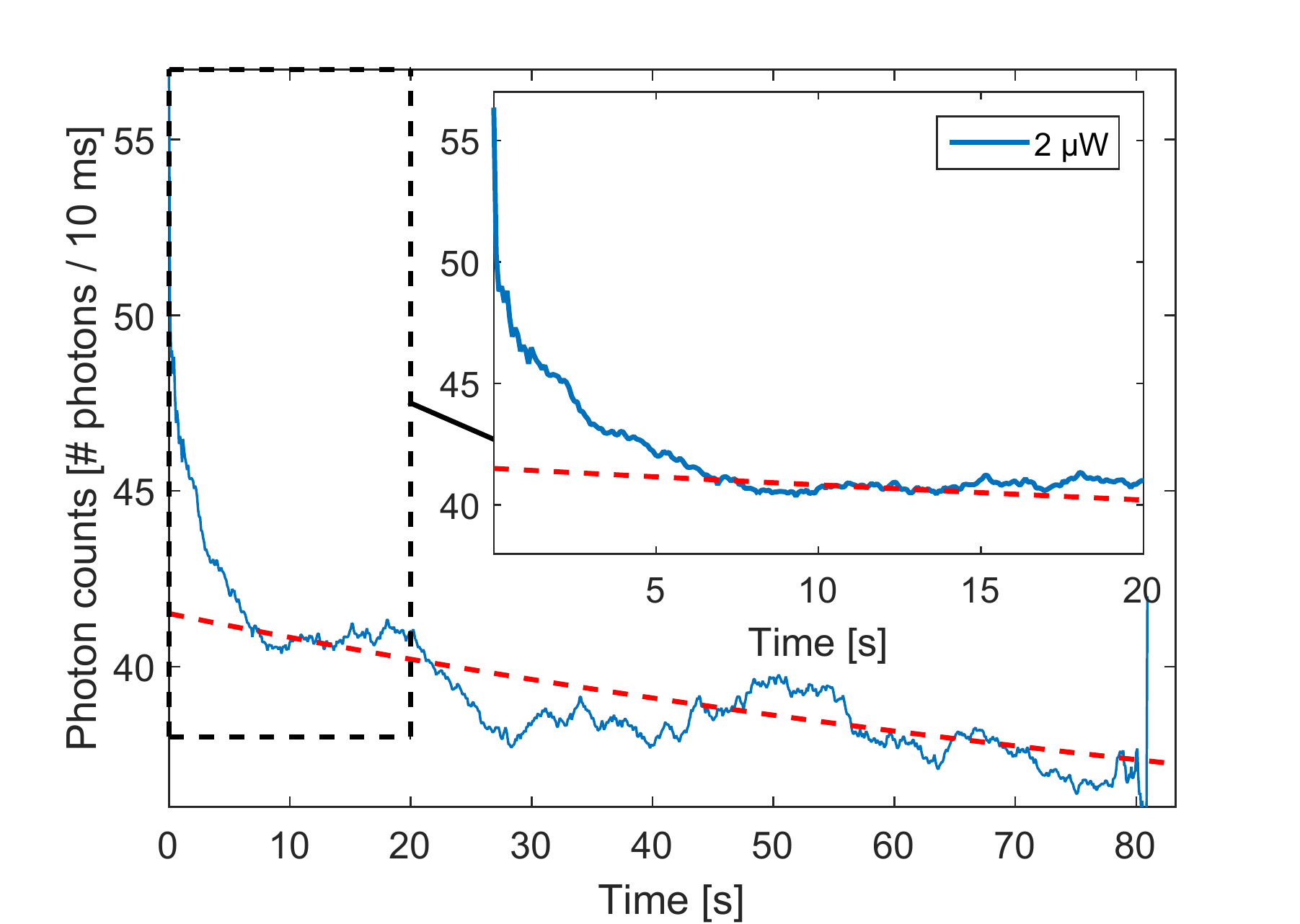}%
\caption{(Color online) Fluorescence decrease for an excitation power of 2 $\mu$W. The simulation result can be seen in the dashed red curve. \label{fig:2uW}}
\end{figure}

\begin{figure}
\includegraphics[width=0.48\textwidth]{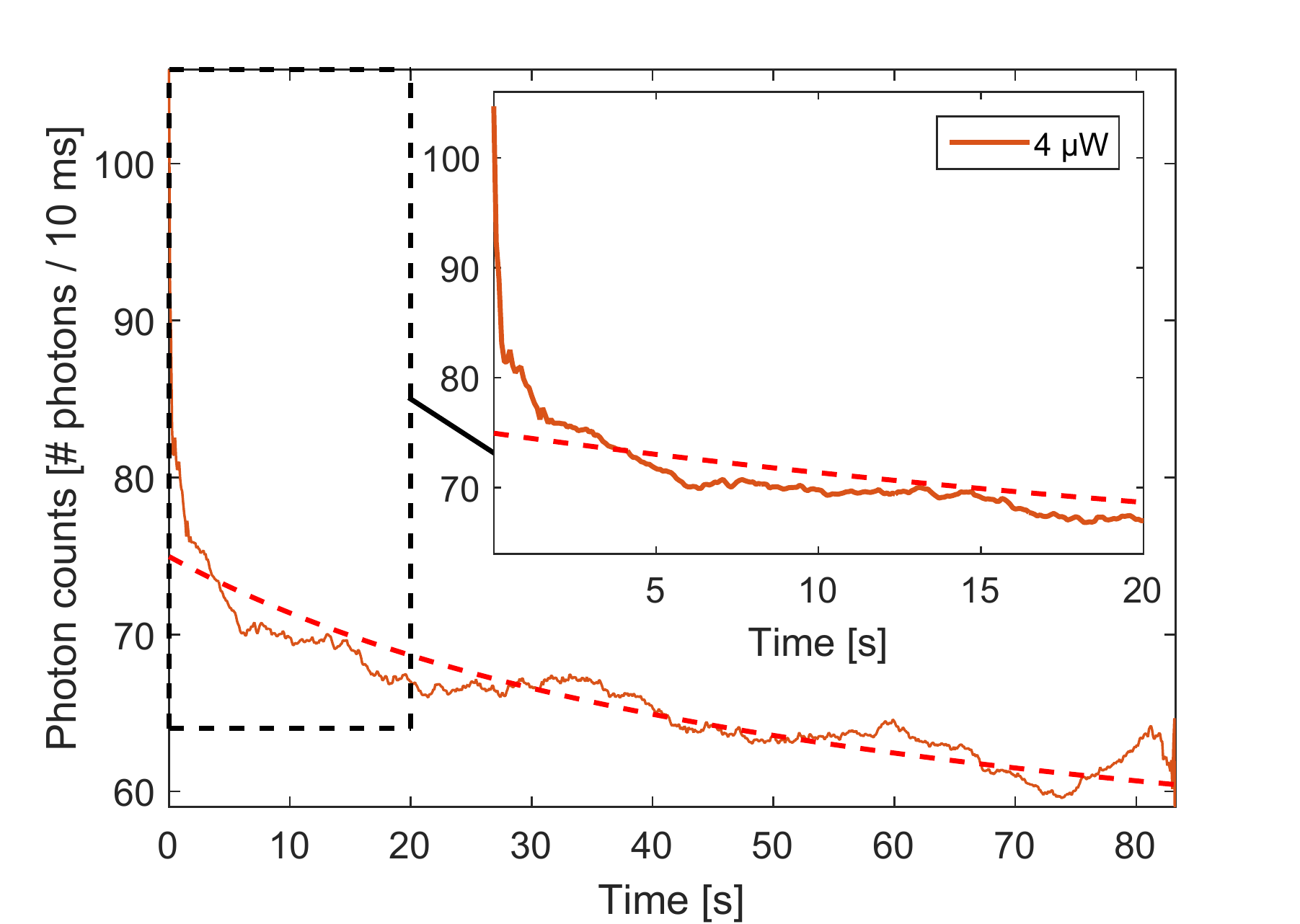}%
\caption{(Color online) Fluorescence decrease for an excitation power of 4 $\mu$W. The simulation result can be seen in the dashed red curve. \label{fig:4uW}}
\end{figure}

\begin{figure}
\includegraphics[width=0.48\textwidth]{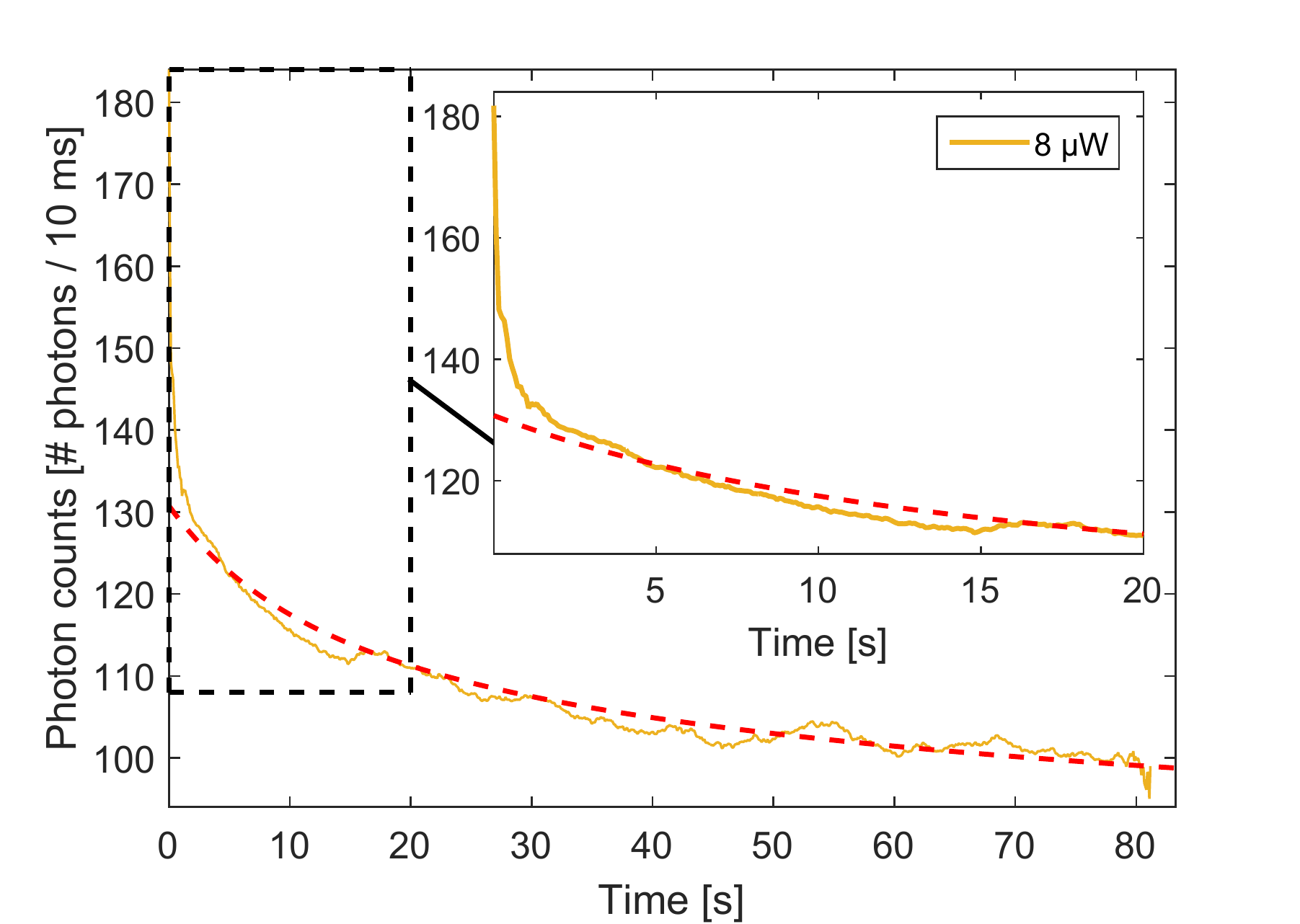}%
\caption{(Color online) Fluorescence decrease for an excitation power of 8 $\mu$W. The simulation result can be seen in the dashed red curve. \label{fig:8uW}}
\end{figure}

\begin{figure}
\includegraphics[width=0.48\textwidth]{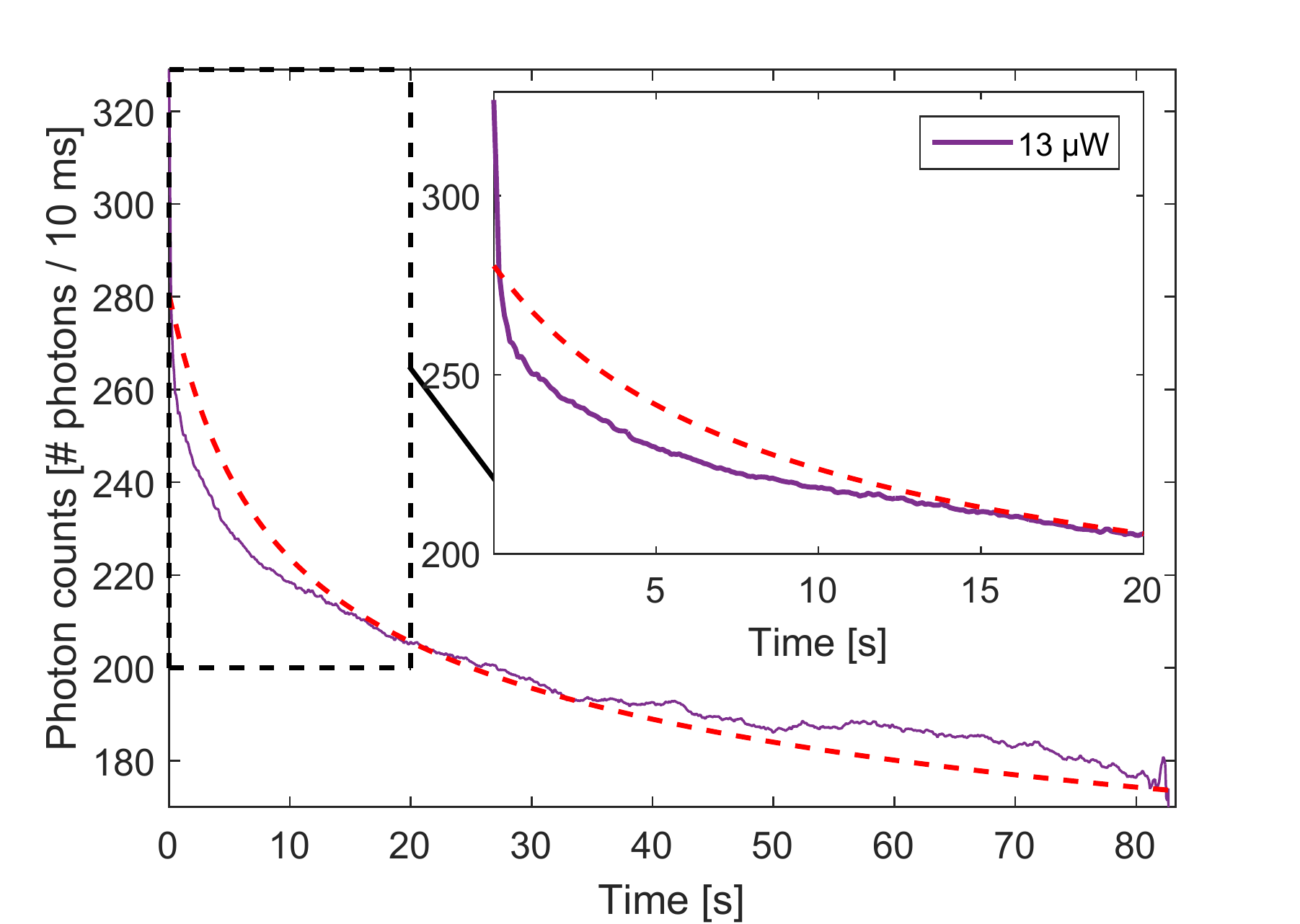}%
\caption{(Color online) Fluorescence decrease for an excitation power of 13 $\mu$W. The simulation result can be seen in the dashed red curve. \label{fig:13uW}}
\end{figure}

\begin{figure}
\includegraphics[width=0.48\textwidth]{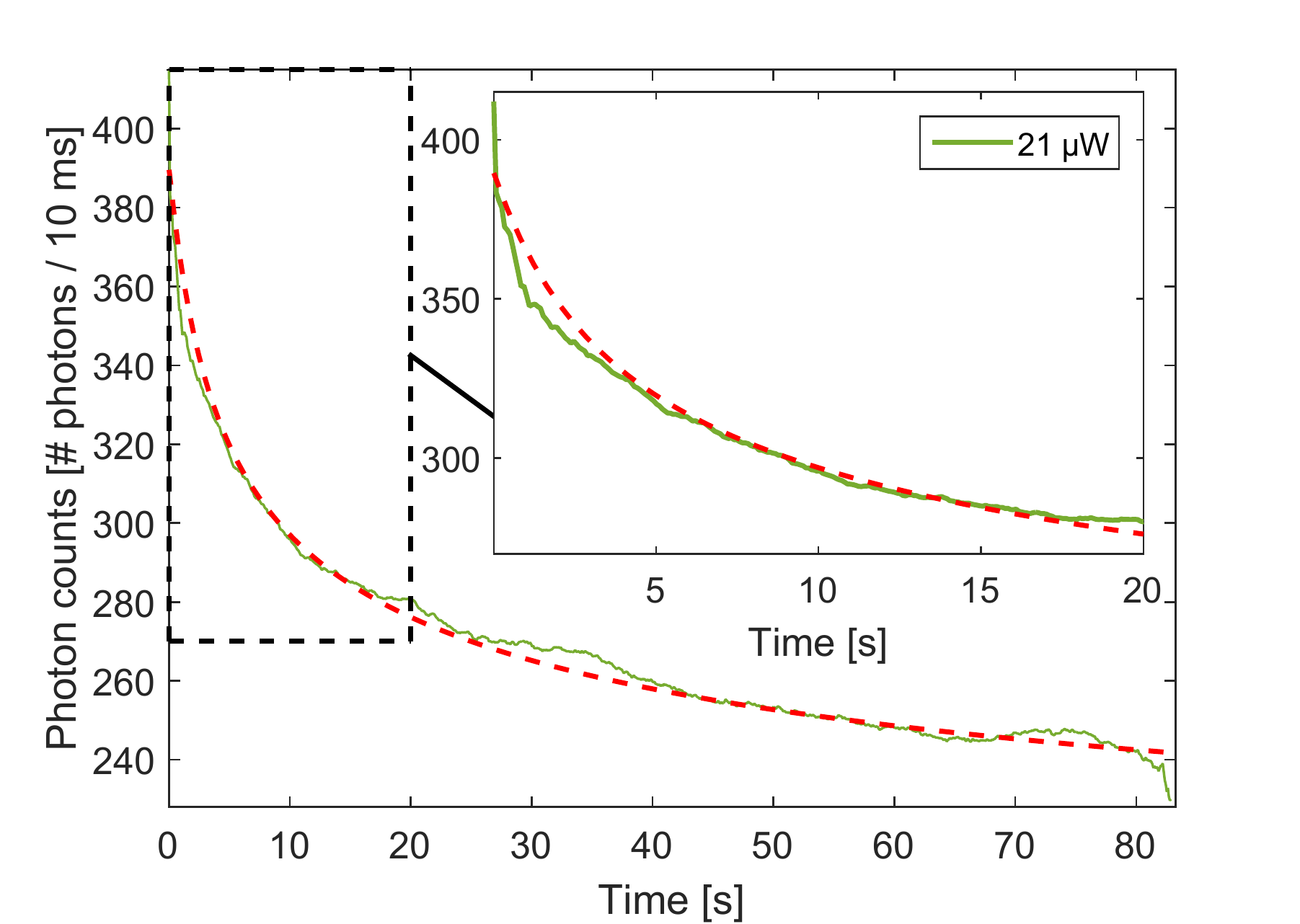}%
\caption{(Color online) Fluorescence decrease for an excitation power of 21 $\mu$W. The simulation result can be seen in the dashed red curve. \label{fig:21uW}}
\end{figure}

\begin{figure}
\includegraphics[width=0.48\textwidth]{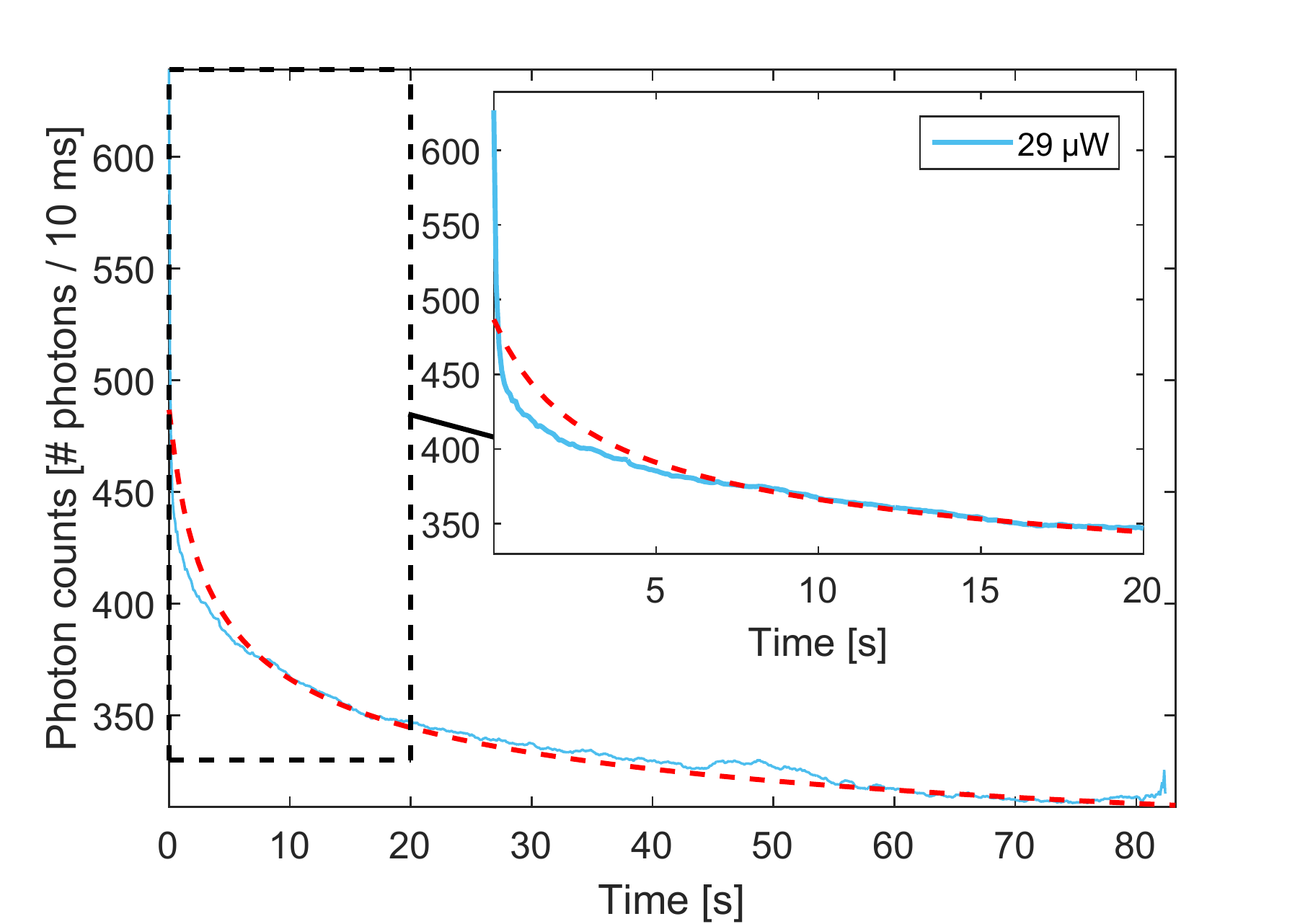}%
\caption{(Color online) Fluorescence decrease for an excitation power of 29 $\mu$W. The simulation result can be seen in the dashed red curve. \label{fig:29uW}}
\end{figure}

\begin{figure}
\includegraphics[width=0.48\textwidth]{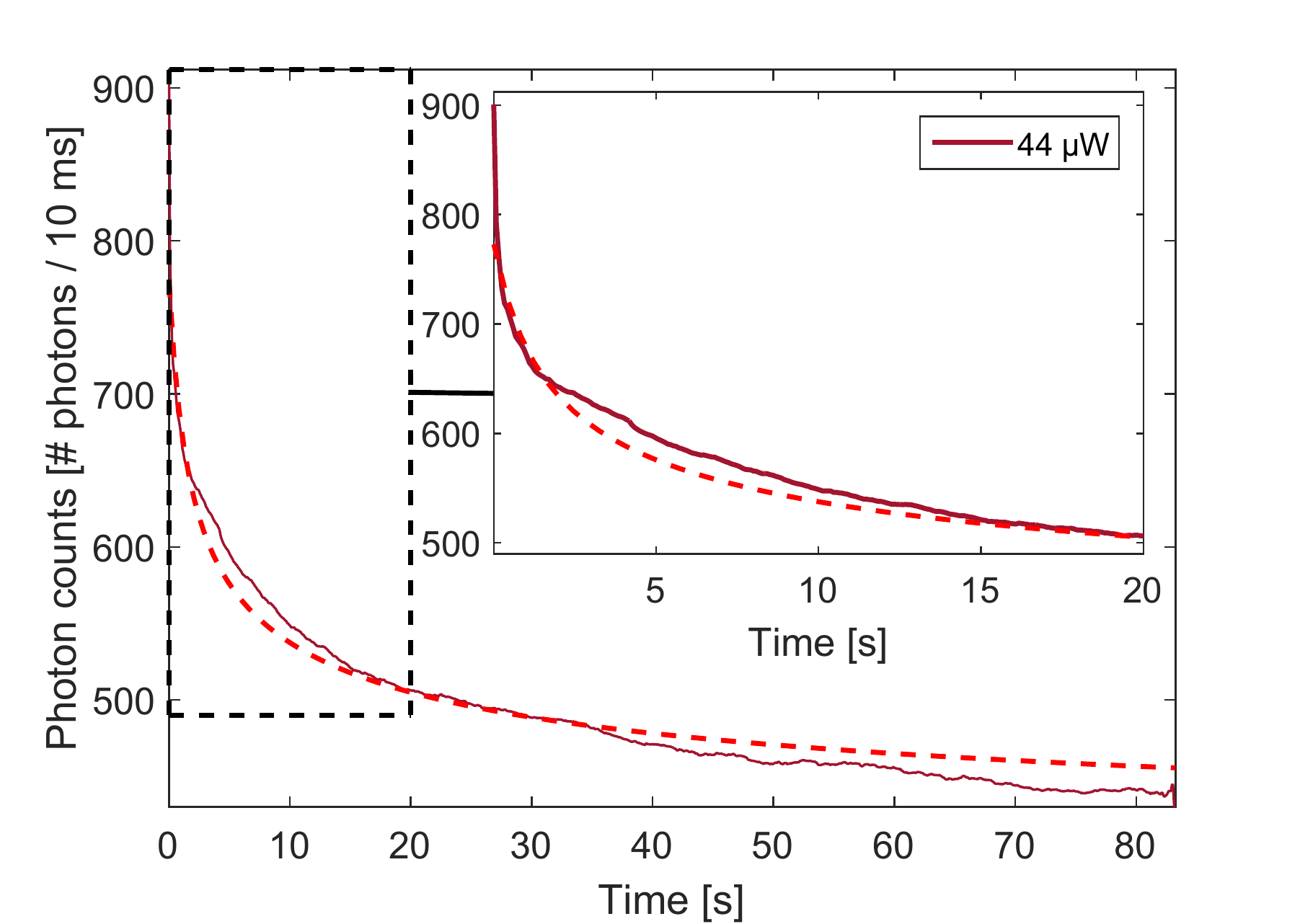}%
\caption{(Color online) Fluorescence decrease for an excitation power of 44 $\mu$W. The simulation result can be seen in the dashed red curve. \label{fig:44uW}}
\end{figure}

The solid curves in figures \ref{fig:2uW}-\ref{fig:44uW} show the fluorescence decrease as a function of time for different input intensities. The dashed red curves are the simulation results from the rate equation model explained above in section \ref{sec:Model}, as well as in section \ref{sec:PermanentTrapping} Permanent Trapping and visualized in figure \ref{fig:levels} of the main article. The signal multiplier parameter is optimized individually for each experimental curve, whilst the initial trap rate and the background fluorescence are optimized globally. This optimization of the parameters was performed to minimize the overall parameter count whilst still accounting for drift in detection efficiency of the setup during the measurements. All experimental curves were taken with an applied magnetic field of 0.2 mT, which in our cryostat minimizes the total magnetic field, as discussed in the main text. The simulation fits are best for the higher power experiments, but seems overall to have difficulties capturing the initial fast decay, which might indicate that something not accounted for in the model is happening in the beginning of the process.

\FloatBarrier

\bibliography{CeBurning_PhysRevB}

\end{document}